\documentclass[12pt]{article}
\usepackage{epsf}
\usepackage[nonstop]{feynmp} 
\usepackage{here}
\usepackage{amsmath}
\def\cM{{\cal M}} 
\def\cO{{\cal O}}
\def\cK{{\cal K}} 
 
\def \be {\begin{equation}}
\def \ee {\end{equation}}
\def \bea {\begin{eqnarray}}
\def \ea {\end{eqnarray}}
\def \ep {\epsilon}
\def \ga {\gamma}

\def \la {\lambda}
\def \no {\nonumber}

\def \dd {{\rm d}}

\def \ps {p\hspace{-0.43em}/}
\def \mps {p\hspace{-0.45em}/}
\def \ns {n\hspace{-0.51em}/}
\def \mns {n\hspace{-0.53em}/}

\newcommand{\To}[2]{\stackrel{#1}{\hbox to #2 pt{\rightarrowfill}}}

\def \pa {\partial}
\def \c {\hspace{-0.2em} \cdot}
\def \ni {\noindent}

\def\np#1#2#3{{\sl  Nucl.~Phys.\/}~{\bf B#1} {(#2) #3}}
\def\spj#1#2#3{{\sl Sov.~Phys.~JETP\/}~{\bf #1} {(#2) #3}}
\def\plb#1#2#3{{\sl Phys.~Lett.\/}~{\bf B#1} {(#2) #3}}
\def\pl#1#2#3{{\sl Phys.~Lett.\/}~{\bf #1} {(#2) #3}}
\def\prd#1#2#3{{\sl Phys.~Rev.\/}~{\bf D#1} {(#2) #3}}
\def\pr#1#2#3{{\sl Phys.~Rep.\/}~{\bf #1} {(#2) #3}}

\begin{document}  
\vspace*{-2cm}  
\renewcommand{\thefootnote}{\fnsymbol{footnote}}  
\begin{flushright}  
hep-ph/0005316\\
DTP/00/18\\  
\end{flushright}  
\vskip 65pt  
\begin{center}  
{\Large \bf New Insights into the Perturbative Structure\\[3truemm] 
            of Electroweak Sudakov Logarithms}\\
    
\vspace{1.2cm} 
{\bf  
W. Beenakker\footnote{W.J.P.Beenakker@durham.ac.uk}%
            \footnote{Supported by a PPARC Research Fellowship}  
and
A. Werthenbach\footnote{Anja.Werthenbach@durham.ac.uk}%
              \footnote{Supported by a DAAD Doktorandenstipendium (HSP III)} 
}\\  
\vspace{20pt}  
{\sf Department of Physics, University of Durham,  Durham DH1 3LE, U.K.\\ }

\vspace{70pt}  
\begin{abstract}
To match the expected experimental precision at future linear colliders, 
improved theoretical predictions beyond next-to-leading order are required. 
At the anticipated energy scale of $\sqrt{s}=1$~TeV the electroweak virtual 
corrections are strongly enhanced by collinear-soft Sudakov logarithms of the 
form $\log^2(s/M^2)$, with $M$ being the generic mass scale of the $W$ 
and $Z$ bosons. By choosing an appropriate gauge, we have developed a formalism
to calculate such corrections for arbitrary electroweak processes. 
As an example we consider in this letter the process $e^+e^- \to f \bar{f}$
and study the perturbative structure of the electroweak Sudakov logarithms by
means of an explicit two-loop calculation. In this way we investigate how the
Standard Model, with its mass gap between the photon and $Z$ boson in the
neutral sector, compares to unbroken theories like QED and QCD. 
In contrast to what is known for unbroken theories we find that the Sudakov
logarithms are not exclusively given by the so-called rainbow diagrams, owing
to the mass gap and the charged-current interactions. In spite of this, we
nevertheless observe that the two-loop corrections are consistent with an
exponentiation of the one-loop corrections. In this sense the Standard Model
behaves like an unbroken theory at high energies.  

\end{abstract}
\end{center}  
\vskip12pt

\setcounter{footnote}{0}  
\renewcommand{\thefootnote}{\arabic{footnote}}  
  
\vfill  
\clearpage  
\setcounter{page}{1}  
\pagestyle{plain} 
\section{Introduction} 
At the next generation of colliders center-of-mass energies will be reached 
that largely exceed the electroweak scale. For instance, the energy at a future
linear $e^+e^-$ collider is expected to be in the TeV range~\cite{LCreport}.
At these energies one enters the realm of large perturbative corrections.
Even the effects arising from weak corrections are expected to be of the order 
of $10\%$ or more~\cite{size,wwhea}, i.e.\ just as large as the well-known
electromagnetic corrections. In order not to jeopardize any of the 
high-precision studies at these high-energy colliders, it is therefore 
indispensable to improve the theoretical understanding of the radiative 
corrections in the weak sector of the Standard Model (SM). In particular this 
will involve a careful analysis of effects beyond first order in the 
perturbative expansion in the (electromagnetic) coupling $\alpha=e^2/(4\pi)$.\\

\ni The dominant source of radiative corrections at TeV-scale energies 
is given by logarithmically enhanced effects of the form 
$\alpha^n\log^m(M^2/s)$ for $m\le 2n$, involving particle masses $M$ well 
below the collider energy $\sqrt{s}$. A natural way of controlling the
theoretical uncertainties would therefore consist in a comprehensive study of 
these large logarithms, taking into account all possible sources (i.e.\
ultraviolet, soft, and collinear). In first approximation the so-called Sudakov
logarithms $\propto \alpha^n\log^{2n}(M^2/s)$, arising from collinear-soft 
singularities~\cite{sudakov}, constitute the leading contribution to the large 
electroweak correction factors. Recent studies have focused on these Sudakov 
effects in the process $e^+e^- \to f\bar{f}$ \cite{kuehn1}--\cite{melles}. 
Unfortunately the three independent studies are in mutual disagreement, 
exhibiting strikingly different higher-order results already for the virtual 
corrections. The main cause for the differences can be traced back to the use 
of different assumptions concerning the exponentiation properties of the 
Sudakov logarithms in the SM. Many of these assumptions are based on the 
analogy with unbroken theories like QED and QCD, where the resummation of the 
higher-order effects amounts to an exponentiation of the one-loop corrections 
(see for instance Refs.~\cite{sudakov} and \cite{expon}--\cite{frenkel}). 
However, the SM is a broken gauge theory with a large mass gap in the neutral 
sector between the massless photon and the massive $Z$ boson, making it a
theory with more than one mass scale. As such it 
remains to be seen how much of the analogy with unbroken theories actually 
pertains to the SM.\\ 

\ni In this letter we therefore focus on the virtual Sudakov logarithms 
in the reaction $e^+e^- \to f \bar{f}$, in an attempt to clarify how the 
$\cO(\alpha^2)$ effects relate to the $\cO(\alpha)$ ones. In this
way we identify to what extent the SM behaves like an unbroken theory 
at high energies. Moreover, since the Sudakov logarithms originate from the
exchange of soft, effectively on-shell gauge bosons, many of the features 
derived for the virtual corrections are intimately related to properties of 
the corresponding real-gauge-boson emission processes.

\section{Electroweak Sudakov logarithms in \boldmath $e^+e^- \to f \bar{f}$}

In order to facilitate the calculation of the one- and two-loop Sudakov 
logarithms, we work in the Coulomb gauge for both massless and massive gauge
bosons (the subtleties associated with the Coulomb gauge for massive gauge 
bosons will be discussed in Ref.~\cite{coulomb}). Working in this special 
gauge has the advantage that all virtual Sudakov logarithms are contained 
exclusively in the self-energies of the external on-shell 
particles~\cite{taylor,coulomb} or the self-energies of any intermediate 
particle that happens to be effectively on-shell.%
\footnote{Note that similar simplifications can probably be obtained 
          equally well by working in an axial gauge, see for instance
          Ref.~\cite{frenkel} for massless particles.}
The latter is, for instance, needed for the production of near-resonance 
unstable particles. The elegance of this method lies in its universal nature, 
with the Sudakov logarithms originating from vertex, box etc.\ corrections 
being suppressed. Once all self-energies to all on-shell/on-resonance SM 
particles have been calculated, the prediction of the Sudakov form factor for 
an {\it arbitrary} electroweak process becomes trivial. The relevant 
self-energies for the calculation of the Sudakov logarithms involve the 
exchange of collinear-soft gauge bosons, including their potential mixing
 with  the corresponding would-be Goldstone bosons. The collinear-soft
  exchange of 
fermions and ghosts leads to suppressed contributions, since the
propagators of these particles do not have the required pole structure.

\subsection{The fermionic self-energy at one-loop level}

\ni As mentioned above, in order to determine the Sudakov logarithms in 
the process $e^+e^- \to f \bar{f}$ one has to calculate the external 
self-energies (i.e.\ the wave-function factors) of all fermions involved in the
process.\\

\ni Consider to this end the fermionic one-loop self-energy 
$\Sigma^{\,(1)}(p,n,M_1)$, originating from the emission of a gauge boson 
$V_1$ with loop-momentum $k_1$ and mass $M_1$ from a fermion $f$ with momentum 
$p$:\\
\begin{minipage}[c]{15.5cm}
\vspace{-1.3cm}
\begin{figure}[H]
\hspace{7cm}
  \begin{fmffile}{self}
  \begin{fmfgraph*}(140,140) \fmfpen{thin} \fmfleft{i1} \fmfright{o1}
  \fmf{fermion,tension=0.7,label=$f(p)$,l.side=right,width=0.6}{i1,v1} 
  \fmf{fermion,tension=0.5,label=$f_1(p\!-\!k_1)$,l.side=right,width=0.6}
      {v1,v3} 
  \fmf{fermion,tension=0.7,label=$f(p)$,l.side=right,width=0.6}{v3,o1} 
  \fmfdot{v1} \fmfdot{v3} 
  \fmf{boson,left,tension=-1,label=$V_1(k_1)$,l.side=left,width=0.6}{v1,v3}
  \fmf{phantom,left}{v1,v3}
  \fmffreeze
  \fmf{phantom_arrow}{v1,v3}
  \end{fmfgraph*}
  \end{fmffile}
\end{figure}
{}\vspace*{-4.3cm}
\bea
\hspace{-5cm}  -\,i\,\, \Sigma^{\,(1)}(p,n,M_1)\,\,\,\,\, =\,   \no
\ea
{}\vspace*{0.2cm}
\end{minipage}\\
\ni Here $n$ is the unit vector in the time direction, which enters by virtue 
of using the Coulomb gauge. In the high-energy limit the fermion mass in the 
numerator of the fermion propagator can be neglected and similarly the
contribution involving a mixed gauge-boson -- Goldstone-boson propagator can
be discarded. The self-energy 
$\Sigma^{\,(1)}$ then contains an odd number of $\ga$-matrices, leading to 
the following natural decomposition in terms of the two possible structures 
$\ps$ and $\ns$:
\bea
  \Sigma^{\,(1)}(p,n,M_1) &\approx& 
      \Bigl[ \,\mps\,\Sigma_p^{\,(1)}(n\,\c p,p^2,M_1)
           + \mns\:\frac{p^2}{n\,\c p}\,\Sigma_n^{\,(1)}(n\,\c p,p^2,M_1)
      \Bigr] e^2\,\Gamma_{\!\!ff_1\!V_1}^{\,^{\scriptstyle 2}}~.
\ea
The coupling factor $\Gamma_{\!\!ff_1\!V_1}$ is defined according to
\bea
   \Gamma_{\!\!ff_1\!V_1} &=& V_{\!\!ff_1\!V_1} - \ga_5\,A_{\!ff_1\!V_1}~,
\ea
where $V_{\!\!ff_1\!V_1}$ and $A_{\!ff_1\!V_1}$ are the vector and axial-vector
couplings of the fermion $f$ to the exchanged gauge boson $V_1$. In our 
convention these coupling factors read
\bea
 \Gamma_{\!\!ff\,\ga} &=& -\,Q_f, \qquad
 \Gamma_{\!\!ffZ} = \frac{(1-\ga_5)\,I_f^3 - 2\,Q_f\,\sin^2\theta_{\rm w}}
                         {2\cos\theta_{\rm w}\,\sin\theta_{\rm w}}, \qquad
 \Gamma_{\!\!ff'\,\!W} = \frac{(1-\ga_5)}{2\sqrt{2}\,\sin\theta_{\rm w}}~.
\ea 
Here $I_f^3$ is the third component of the weak isospin, $e\,Q_f$ is the
electromagnetic charge, and $\theta_{\rm w}$ is the weak mixing angle.\\

\ni The contribution to the external wave-function factor now amounts to
multiplying the self-energy by $\,i/\mps\,$ on the side where it is 
attached to the rest of the scattering diagram and by the appropriate fermion 
source on the other side. Finally the square root should be taken of the
external wave-function factor, i.e.\ the one-loop contribution should be 
multiplied by the usual factor $1/2$. For an initial-state fermion, 
for example, one obtains%
\footnote{For an outgoing fermion one obtains $\frac{1}{2}\,\bar{u}_f(p)\,
          \bar{\Sigma}_f^{\,(1)}(s,m_f^2,M_1)$, where $\bar{\Sigma}_f^{\,(1)}$
          can be derived from $\Sigma_f^{\,(1)}$ by reversing the sign in front
          of $\ga_5$.}
\bea
  \frac{1}{2}\,\,\frac{i}{\mps}\,\Bigl[ -i\,\Sigma^{\,(1)}(p,n,M_1) \Bigr]\,
  u_f(p) 
  &\approx& \frac{e^2}{2}\,\Gamma_{\!\!ff_1\!V_1}^{\,^{\scriptstyle 2}}\, 
      \Bigl[ \Sigma_p^{\,(1)}(n\,\c p,m_f^2,M_1)
             + 2\,\Sigma_n^{\,(1)}(n\,\c p,m_f^2,M_1)
      \Bigr]\,u_f(p) \no\\[1mm]
                &\equiv& \frac{1}{2}\,\Sigma_f^{\,(1)}(s,m_f^2,M_1)\,u_f(p)~,
\ea
where $m_f$ is the mass of the external fermion and $\sqrt{s}=2\,p_0$ is the 
center-of-mass energy of the process $e^+e^- \to f \bar{f}$. Therefore the 
quantity of interest is actually $\Sigma_f^{\,(1)}$, which can be extracted 
from the full fermionic self-energy by means of the projection
\bea
  \Sigma_f^{\,(1)}(s,m_f^2,M_1) &=& \frac{i}{2\,p_0}\,\bar{u}_f(p)\,
              \biggl\{ \frac{\pa}{\pa p_0}\,\Bigl[ -i\,\Sigma^{\,(1)}(p,n,M_1)
                                            \Bigr] \biggr\}\,u_f(p) \no\\[2mm]
                                &\approx& -\,e^2\,
              \Gamma_{\!\!ff_1\!V_1}^{\,^{\scriptstyle 2}}\, \int 
              \frac{\dd^4 k_1}{(2\,\pi)^4}\,
              \frac{4\,p_{\mu}\,p_{\nu}}{[(p-k_1)^2-m_{f_1}^2+i\epsilon]\,^2}\,
              P^{\mu\nu}(k_1,n)~,
\ea
containing the gauge-boson propagator in the Coulomb gauge
\bea
  P^{\mu\nu}(k_1,n) &=& \frac{-i}{k_1^2-M_1^2+i \epsilon}\,\biggl[ g^{\mu\nu}
     + \frac{k_1^{\mu}k_1^{\nu} - k_1^0\,( k_1^{\mu}n^{\nu}+n^{\mu}k_1^{\nu} )}
            {\vec{k}_1^{\,2}} \biggr]~.
\ea
Note that the loop-momentum $k_1$ has been neglected in the numerator of the
fermion propagator, since only collinear-soft gauge-boson momenta will give 
rise to the Sudakov logarithms. The mass of the fermion inside the loop, 
$m_{f_1}$, is at best of the order of the $Z$-boson mass (for the top-quark). 
At the leading-logarithmic level it therefore only enters as an independent 
mass scale if the exchanged gauge boson is a photon (i.e.\ $m_{f_1}=m_f$), 
where it is needed for the regularization of collinear singularities. 
Therefore we restrict the arguments of $ \Sigma_f^{\,(1)}$ to the energy and 
mass of the emitting fermion and the mass of the exchanged gauge boson, which 
are the three relevant scales for describing the Sudakov logarithms.\\

\ni Having two canonical momenta at our disposal, i.e.\ $p$ and
$n$, we define the following Sudakov parametrisation of the gauge-boson 
loop-momentum $k_1$:
\bea
  k_1= v_1\,q + u_1\,\bar{q} + k_{1_\bot}~,
\ea
with 
\bea
  p^{\mu} \equiv (E,\beta_f E,0,0)~, 
  & & \qquad\qquad \beta_f = \sqrt{1-m_f^2/E^2}~, 
      \qquad\qquad s=4\, E^2, \no \\[1mm] 
  q^\mu = (E,E,0,0)~,                
  & & \qquad\qquad \bar{q}^\mu = (E,- E,0,0)~,
      \qquad\qquad k_{1_\bot}^{\mu} = (0,0,\vec{k}_{1_\bot})~.
\ea
In terms of this parametrisation, the integration measure $\dd^4 k_1$,
the invariants $(p\, \c k_1)$ and $k_1^2$, and the gauge-boson energy $k_1^0$
read   
\bea
  \dd^4 k_1    &=& \pi\,\frac{s}{2}\,\dd v_1\,\dd u_1\, 
                   \dd \vec{k}_{1_\bot}^{\,2}~, \no\\[1mm]
  (p\, \c k_1) &=& \frac{s}{4}\, [\,v_1\,(1-\beta_f) + u_1\,(1+\beta_f)\,]
                   \approx \ 
                   \frac{s}{2}\,\Bigl( u_1+\frac{m_f^2}{s}\,v_1 \Bigr)~,
                   \no\\[2mm]
  k_1^2        &=& s\,v_1\,u_1 - \vec{k}_{1_\bot}^{\,2}
                   \qquad {\rm and} \qquad 
                   k_1^0 \ =\ \frac{\sqrt{s}}{2}\,(v_1+u_1)~.
\ea
The term containing the fermion mass $m_f$ is needed for the exchange of 
photons only, regulating the collinear singularity at $u_1=0$. For the exchange
of a massive gauge boson the mass $M_1$ will be the dominant collinear as well
as infrared regulator.\\  

\ni The $v_1$-integration is restricted to the interval $\,0\le v_1 \le 1$, as 
a result of the requirement of having poles in both hemispheres of the complex 
$u_1$-plane. The residue is then taken in the lower hemisphere in the pole of
the gauge-boson propagator: $s\,v_1\,u_1^{\rm res} = \vec{k}_{1_\bot}^{\,2} 
+ M_1^2 \equiv s\,v_1\,y_1$. Finally, $\vec{k}_{1_\bot}^{\,2}$ is substituted 
by $y_1$, with the condition $\,\vec{k}_{1_\bot}^{\,2} \ge 0\,$ translating 
into $\,v_1\,y_1 \ge M_1^2/s$. The one-loop Sudakov contribution to 
$\Sigma_f^{\,(1)}$ now reads 
\bea
\label{kernel}
  \Sigma^{\,(1)}_f(s,m_f^2,M_1) &\approx& -\,\frac{\alpha}{\pi}\,
     \Gamma_{\!\!ff_1\!V_1}^{\,^{\scriptstyle 2}}\,\int_0^{\infty} \dd y_1 \,
     \int_0^1 \dd v_1 \ \frac{\Theta(v_1 y_1-\frac{M_1^2}{s})}
                              {(y_1+\frac{m_f^2}{s}v_1)\,(v_1+y_1)} 
     \no \\[1mm]
                                &\approx& -\,\frac{\alpha}{\pi}\,
     \Gamma_{\!\!ff_1\!V_1}^{\,^{\scriptstyle 2}}\,\int_0^1 \frac{\dd y_1}{y_1}
     \int_{y_1}^1 \frac{\dd z_1}{z_1}\ \cK^{\,(1)}(s,m_f^2,M_1,y_1,z_1)~,
\ea
with the integration kernel $\cK^{\,(1)}$ given by 
\bea
  \cK^{\,(1)}(s,m_f^2,M_1,y_1,z_1) &=& 
           \Theta\Bigl( y_1 z_1-\frac{M_1^2}{s} \Bigr)\,
           \Theta\Bigl (y_1-\frac{m_f^2}{s}\,z_1\Bigr)~.
\ea
Here we introduced the energy variable $\,z_1 = v_1 + y_1\,$ and made use of 
the fact that only collinear-soft gauge-boson momenta are responsible for the 
quadratic large-logarithmic effects: $y_1,\,z_1 \ll 1$. As a result, the 
gauge boson inside the loop is effectively on-shell and transversely 
polarized.\\

\ni The exchanged gauge boson can either be a massless photon ($\ga$) or one 
of the massive weak bosons ($W$ or $Z$). The associated mass gap gives rise to 
distinctive differences in the two types of contributions 
(see Sect.~\ref{sec:integrals}). Bearing in mind that the SM is not parity
conserving, we sum over the gauge-boson contributions and present the full 
one-loop Sudakov correction factor for right- and left-handed 
fermions/antifermions separately:
\begin{subequations} 
\label{delta1}
\bea
\label{delta1+}
  \delta^{\,(1)}_{f_R} &=& \delta^{\,(1)}_{\bar{f}_L} \ =\   
        \left( \frac{Y_f^R}{2\cos\theta_{\rm w}} \right)^{\!2} \!{\rm L}(M,M)
        + Q_f^2\,\bigg[ {\rm L_{\ga}}(\la,m_f)-{\rm L}(M,M) \bigg]~, 
        \\
\label{delta1-}
  \delta^{\,(1)}_{f_L} &=& \delta^{\,(1)}_{\bar{f}_R} \ =\ 
        \left[ \frac{C_F}{\sin^2\theta_{\rm w}} 
        + \left( \frac{Y_f^L}{2\cos\theta_{\rm w}} \right)^{\!2}\, \right]\!  
          {\rm L}(M,M)
        + Q_f^2 \,\bigg[ {\rm L_{\ga}}(\la,m_f)- {\rm L}(M,M) \bigg]~,
\ea
\end{subequations} 
with 
\bea
\label{L}
 {\rm L}(M_1,M_2)              &=& -\,\frac{\alpha}{4\,\pi}\,
          \log\left( \frac{M_1^2}{s} \right) 
          \log\left( \frac{M_2^2}{s} \right)~, \\[1mm]
\label{Lgamma}
 {\rm L_{\ga}}(\la,M_1) &=& -\,\frac{\alpha}{4\,\pi}\, 
          \left[ \log^{\, 2}\!\left( \frac{\la^2}{s} \right) 
               - \log^{\, 2}\!\left( \frac{\la^2}{M_1^2} \right) 
          \right]~.
\ea
Note that these correction factors are the same for incoming as well as
outgoing particles. In Eq.~(\ref{delta1}) $Y_f^{R,L}$ denotes the right- and 
left-handed hypercharge of the external fermion, which is connected to the 
third component of the weak isospin $I^3_f$ and the electromagnetic charge 
$e\,Q_f$ through the Gell-Mann -- Nishijima relation 
$Q_f = I^3_f + Y_f^{R,L}/2$. 
The coefficient $C_F=3/4\,$ is the Casimir operator in the fundamental 
representation of $SU(2)$ and $\la$ is the fictitious (infinitesimally small) 
mass of the photon needed for regularizing the infrared singularity at $z_1=0$.
For the sake of calculating the leading Sudakov logarithms, the masses of the 
$W$ and $Z$ bosons can be represented by one generic mass scale $M$. Note that 
the terms proportional to $Q_f^2$ in Eq.~(\ref{delta1}) are the result of the 
mass gap between the photon and the weak bosons. \\

\ni In the process $e^+e^- \to f \bar{f}\,$ the one-loop correction 
factors presented in Eq.~(\ref{delta1}) contribute in the following way to the 
polarized matrix element, bearing in mind that at high energies the helicity 
eigenstates are equivalent to the chiral eigenstates:
\bea
 \cM_{e^+_R e^-_L \to f_L \bar{f}_R}^{\rm 1-loop,\,\,sudakov} &=&
      \frac{1}{2}\,\Bigl[ \delta^{\,(1)}_{e^+_R} + \delta^{\,(1)}_{e^-_L}
                        + \delta^{\,(1)}_{f_L} + \delta^{\,(1)}_{\bar{f}_R} 
                   \Bigr] \cM_{e^+_R e^-_L \to f_L \bar{f}_R}^{\rm born}~,
\ea
and similar expressions for the other possible helicity combinations.

\subsection{The fermionic self-energy at two-loop level}

\ni At two-loop accuracy one has to take the following five generic 
sets of diagrams into account: \vspace*{2mm}\\
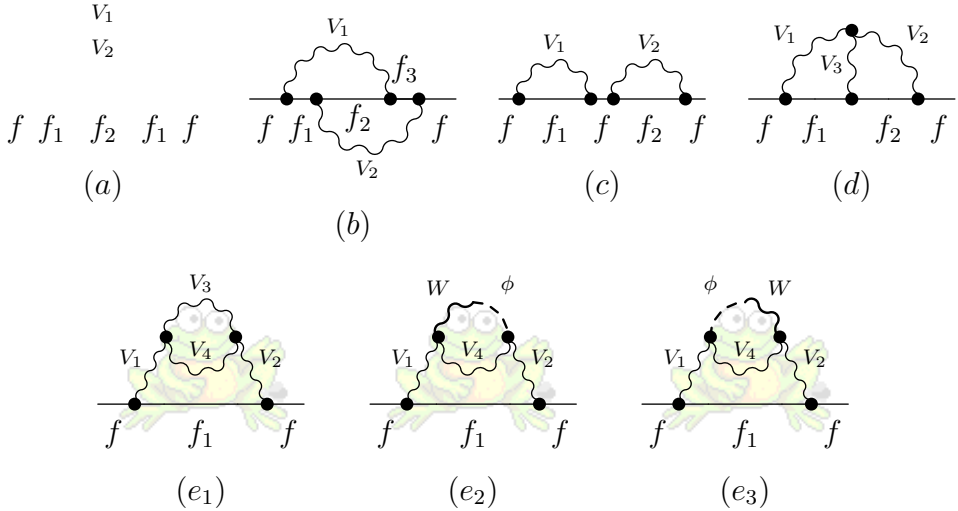
\begin{figure}[H]
\vspace*{3.2cm}

\hspace*{3cm}{\epsfysize=1.65cm{\epsffile{ frog4.epsi}}}
\vspace*{-2cm}
\hspace*{1.3cm}{\epsfysize=1.65cm{\epsffile{ frog4.epsi}}}
\vspace*{-1.65cm}
\hspace*{1.4cm}{\epsfysize=1.65cm{\epsffile{ frog4.epsi}}}
\vspace*{-1.7cm}

\begin{fmffile}{2l}
\hspace*{1.2cm}
\begin{fmfgraph*}(78,52) \fmfpen{thin} \fmfleft{i1} \fmfright{o1}
\fmf{phantom,tension=1}{i1,v1,v2,v5,v3,v4,o1}
\fmf{plain,tension=1,label=$f$,l.side=right,width=0.6}{i1,v1} 
\fmf{plain,tension=1.5,label=$f_1$,l.side=right,width=0.6}{v1,v2}
\fmf{plain,tension=0.5,label=$f_2$,l.side=right,width=0.6}{v2,v3}
\fmf{plain,tension=1.5,label=$f_1$,l.side=right,width=0.6}{v3,v4}
\fmf{plain,tension=1,label=$f$,l.side=right,width=0.6}{v4,o1}
\fmffreeze 
\fmf{boson,left,tension=-1,label=${\scriptstyle
V_1}$,l.side=left,l.dist=4,
     width=0.6}{v1,v4}
\fmf{boson,left,tension=-2,label=${\scriptstyle
V_2}$,l.side=left,l.dist=2,
     width=0.6}{v2,v3}
\fmfdot{v1} \fmfdot{v2} \fmfdot{v3} \fmfdot{v4} 
\fmfv{label=$(a)$,l.dist=27,l.a=-90}{v5}
\end{fmfgraph*}
\hspace{0.3cm}
\begin{fmfgraph*}(78,52) \fmfpen{thin} \fmfleft{i1} \fmfright{o1}
\fmf{phantom,tension=1}{i1,v1,v2,v5,v3,v4,o1}
\fmf{plain,tension=1,label=$f$,l.side=right,width=0.6}{i1,v1} 
\fmf{plain,tension=1.5,label=$f_1$,l.side=right,width=0.6}{v1,v2}
\fmf{plain,tension=0.5,width=0.6}{v2,v3}
\fmfv{label=$f_2$,l.dist=19,l.a=-157.55}{v4}
\fmf{plain,tension=0.1,width=0.6}{v3,v4}
\fmf{plain,tension=1.5,label=$f_3$,l.side=left,width=0.6}{v3,v4}
\fmf{plain,tension=1,label=$f$,l.side=right,width=0.6}{v4,o1}
\fmffreeze 
\fmf{boson,left,tension=-1.5,label=${\scriptstyle
V_1}$,l.side=left,l.dist=4,
     width=0.6}{v1,v3}
\fmf{boson,right,tension=-1.5,label=${\scriptstyle
V_2}$,l.side=right,l.dist=4,
     width=0.6}{v2,v4}
\fmfdot{v1} \fmfdot{v2} \fmfdot{v3} \fmfdot{v4} 
\fmfv{label=$(b)$,l.dist=42,l.a=-90}{v5}
\end{fmfgraph*}
\hspace{0.3cm}
\begin{fmfgraph*}(78,52) \fmfpen{thin} \fmfleft{i1} \fmfright{o1}
\fmf{phantom,tension=1}{i1,v1,v2,v5,v3,v4,o1}
\fmf{plain,tension=3,label=$f$,l.side=right,width=0.6}{i1,v1} 
\fmf{plain,tension=0.1,label=$f_1$,l.side=right,width=0.6}{v1,v2}
\fmf{plain,tension=3,label=$f$,l.side=right,width=0.6}{v2,v3}
\fmf{plain,tension=0.1,label=$f_2$,l.side=right,width=0.6}{v3,v4}
\fmf{plain,tension=3,label=$f$,l.side=right,width=0.6}{v4,o1}
\fmffreeze 
\fmf{boson,left,tension=1,label=${\scriptstyle V_1}$,l.side=left,l.dist=4, 
     width=0.6}{v1,v2}
\fmf{boson,left,tension=1,label=${\scriptstyle V_2}$,l.side=left,l.dist=4,
     width=0.6}{v3,v4}
\fmfdot{v1} \fmfdot{v2} \fmfdot{v3} \fmfdot{v4} 
\fmfv{label=$(c)$,l.dist=27,l.a=-90}{v5}
\end{fmfgraph*}
\hspace{0.3cm}
\begin{fmfgraph*}(78,52) \fmfpen{thin} \fmfleft{i1} \fmfright{o1}
\fmf{phantom,tension=1}{i1,v1,v2,v5,v3,v4,o1}
\fmf{plain,tension=1,label=$f$,l.side=right,width=0.6}{i1,v1} 
\fmf{plain,tension=1.5,width=0.6}{v1,v2}
\fmf{plain,tension=0.5,width=0.6}{v2,v3}
\fmf{plain,tension=1.5,width=0.6}{v3,v4}
\fmf{plain,tension=1,label=$f$,l.side=right,width=0.6}{v4,o1}
\fmffreeze
\fmftop{v6}
\fmf{boson,left,tension=-1,width=0.6}{v1,v4}
\fmf{boson,tension=2,label=${\scriptstyle
V_3}$,l.side=left,l.dist=3,width=0.6}
    {v5,v6}
\fmfv{label=${\scriptstyle V_1}$,l.a=90,l.dist=21}{v1}
\fmfv{label=${\scriptstyle V_2}$,l.a=90,l.dist=21}{v4}
\fmfv{label=${f_1}$,l.a=-90,l.dist=6}{v2}
\fmfv{label=${f_2}$,l.a=-90,l.dist=6}{v3}
\fmfv{label=$(d)$,l.dist=27,l.a=-90}{v5}
\fmfdot{v1} \fmfdot{v5} \fmfdot{v4} \fmfdot{v6} 
\end{fmfgraph*}
\vspace*{2.2cm}

\hspace*{2.5cm}
\begin{fmfgraph*}(78,52) \fmfpen{thin} \fmfleft{i1} \fmfright{o1}
\fmf{phantom,tension=1}{i1,v1,v2,v5,v3,v4,o1}
\fmf{plain,tension=1,label=$f$,l.side=right,width=0.6}{i1,v1} 
\fmf{plain,tension=1.5,width=0.6}{v1,v2}
\fmf{plain,tension=0.5,label=$f_1$,l.side=right,width=0.6}{v2,v3}
\fmf{plain,tension=1.5,width=0.6}{v3,v4}
\fmf{plain,tension=1,label=$f$,l.side=right,width=0.6}{v4,o1}
\fmffreeze
\fmftop{v6,v7,v8,v9}
\fmf{boson,tension=-1,label=${\scriptstyle V_1}$,l.side=left,l.dist=3,
     width=0.6}{v1,v7}
\fmf{boson,tension=-1,label=${\scriptstyle V_2}$,l.side=left,l.dist=3,
     width=0.6}{v8,v4}
\fmf{boson,left,tension=2,label=${\scriptstyle V_3}$,l.side=left,l.dist=4,
     width=0.6}{v7,v8}
\fmf{boson,right,tension=2,label=${\scriptstyle
V_4}$,l.side=left,l.dist=5,
     width=0.6}{v7,v8}
\fmfdot{v1} \fmfdot{v7} \fmfdot{v4} \fmfdot{v8} 
\fmfv{label=$(e_1)$,l.dist=27,l.a=-90}{v5}
\end{fmfgraph*}
\hspace{0.6cm}
\begin{fmfgraph*}(78,52) \fmfpen{thin} \fmfleft{i1} \fmfright{o1}
\fmf{phantom,tension=1}{i1,v1,v2,v5,v3,v4,o1}
\fmf{plain,tension=1,label=$f$,l.side=right,width=0.6}{i1,v1} 
\fmf{plain,tension=1.5,width=0.6}{v1,v2}
\fmf{plain,tension=0.5,label=$f_1$,l.side=right,width=0.6}{v2,v3}
\fmf{plain,tension=1.5,width=0.6}{v3,v4}
\fmf{plain,tension=1,label=$f$,l.side=right,width=0.6}{v4,o1}
\fmffreeze
\fmftop{v6,v7,v8,v9}
\fmf{boson,tension=-1,label=${\scriptstyle V_1}$,l.side=left,l.dist=3,
     width=0.6}{v1,v7}
\fmf{boson,tension=-1,label=${\scriptstyle V_2}$,l.side=left,l.dist=3,
     width=0.6}{v8,v4}
\fmf{phantom,left,tension=2,width=0.6,tag=1}{v7,v8}
\fmfposition
\fmfipath{p[]}
\fmfiset{p1}{vpath1(__v7,__v8)}
\fmfi{boson}{subpath (0,length(p1)/2) of p1}
\fmfi{dashes}{subpath (length(p1)/2,length(p1)) of p1}
\fmf{boson,right,tension=2,label=${\scriptstyle
V_4}$,l.side=left,l.dist=5,
     width=0.6}{v7,v8}
\fmfv{label=${\scriptstyle W}$,l.a=90,l.dist=16}{v7}
\fmfv{label=${\scriptstyle \phi}$,l.a=90,l.dist=16}{v8}
\fmfdot{v1} \fmfdot{v7} \fmfdot{v4} \fmfdot{v8} 
\fmfv{label=$(e_2)$,l.dist=27,l.a=-90}{v5}
\end{fmfgraph*}
\hspace*{0.6cm}
\begin{fmfgraph*}(78,52) \fmfpen{thin} \fmfleft{i1} \fmfright{o1}
\fmf{phantom,tension=1}{i1,v1,v2,v5,v3,v4,o1}
\fmf{plain,tension=1,label=$f$,l.side=right,width=0.6}{i1,v1} 
\fmf{plain,tension=1.5,width=0.6}{v1,v2}
\fmf{plain,tension=0.5,label=$f_1$,l.side=right,width=0.6}{v2,v3}
\fmf{plain,tension=1.5,width=0.6}{v3,v4}
\fmf{plain,tension=1,label=$f$,l.side=right,width=0.6}{v4,o1}
\fmffreeze
\fmftop{v6,v7,v8,v9}
\fmf{boson,tension=-1,label=${\scriptstyle V_1}$,l.side=left,l.dist=3,
     width=0.6}{v1,v7}
\fmf{boson,tension=-1,label=${\scriptstyle V_2}$,l.side=left,l.dist=3,
     width=0.6}{v8,v4}
\fmf{phantom,left,tension=2,width=0.6,tag=1}{v7,v8}
\fmfposition
\fmfipath{p[]}
\fmfiset{p1}{vpath1(__v7,__v8)}
\fmfi{dashes}{subpath (0,length(p1)/2) of p1}
\fmfi{boson}{subpath (length(p1)/2,length(p1)) of p1}
\fmf{boson,right,tension=2,label=${\scriptstyle
V_4}$,l.side=left,l.dist=5,
     width=0.6}{v7,v8}
\fmfv{label=${\scriptstyle W}$,l.a=90,l.dist=16}{v8}
\fmfv{label=${\scriptstyle \phi}$,l.a=90,l.dist=16}{v7}
\fmfdot{v1} \fmfdot{v7} \fmfdot{v4} \fmfdot{v8} 
\fmfv{label=$(e_3)$,l.dist=27,l.a=-90}{v5}
\end{fmfgraph*}
\end{fmffile}
\vspace*{2mm} 
\end{figure}
\ni The fermions $f_{i}$ are fixed by the exchanged gauge bosons $V_{i}$.
Various cancellations are going to take place between all these diagrams. 
In unbroken theories like QED and QCD merely the so-called `rainbow'
diagrams of set (a) survive.
The same holds if all gauge bosons of the theory would have
a similar mass. The unique feature of the SM is that it is only partially
broken, with the electromagnetic gauge group $U(1)_{\rm em} \neq U(1)_Y$ 
remaining unbroken. As such three of the four gauge bosons will acquire a mass,
whereas the photon remains massless and will interact with the charged massive
gauge bosons ($W^{\pm}$). As a consequence, merely calculating the `rainbow'
diagrams will {\it not} lead to the correct result, in contrast to what is claimed
in Refs.~\cite{paolo2,melles}. 

\ni The generic two-loop contribution of Sudakov logarithms to 
$\Sigma_f^{\,(2)}$ reads: 
\bea
\label{twoloop}
  \Sigma^{\,(2)}_f &\approx& \Bigl( -\,\frac{\alpha}{\pi} \Bigr)^2\,
    \Gamma_f^{\,(2)}\, \int_0^1 \frac{\dd y_1}{y_1} 
    \int_{y_1}^1\frac{\dd z_1}{z_1} \int_0^1 \frac{\dd y_2}{y_2} 
    \int_{y_2}^1 \frac{\dd z_2}{z_2}\ \cK^{\,(2)}(y_1,z_1,y_2,z_2)~. 
\ea 
For the five different topologies the various products 
$\Gamma_f^{\,(2)}\,{\scriptstyle \times}\,\cK^{\,(2)}$ of coupling factors 
and integration kernels are given by
\bea
  \mbox{set (a):} &&\!\!\!\!\!
                     \Bigl[ \Gamma_{\!\!ff_1\!V_1}^{\,^{\scriptstyle 2}}\,
                            \cK^{\,(1)}(s,m_f^2,M_1,y_1,z_1) \Bigr]\,
                     \Bigl[ \Gamma_{\!\!f_1f_2V_2}^{\,^{\scriptstyle 2}}\,
                            \cK^{\,(1)}(s,m_f^2,M_2,y_2,z_2) \Bigr]\,
                     \Theta(y_2-y_1)~, \no\\[2mm]
  \mbox{set (b):} && \!\!\!\!\!
                     -\,\Gamma_{\!\!ff_1\!V_1}\,\Gamma_{\!\!f_1f_2V_2}\,
                     \Gamma_{\!\!f_2f_3V_1}\,\Gamma_{\!\!ff_3V_2}\,
                     \cK^{\,(1)}(s,m_f^2,M_1,y_1,z_1)\,
                     \cK^{\,(1)}(s,m_f^2,M_2,y_2,z_2)~, \no\\[3mm]
  \mbox{set (c):} && \!\!\!\!\!
                     \Bigl[ \Gamma_{\!\!ff_1\!V_1}^{\,^{\scriptstyle 2}}\,
                            \cK^{\,(1)}(s,m_f^2,M_1,y_1,z_1) \Bigr]\,
                     \Bigl[ \Gamma_{\!\!ff_2V_2}^{\,^{\scriptstyle 2}}\,
                            \cK^{\,(1)}(s,m_f^2,M_2,y_2,z_2) \Bigr]~, 
                     \no\\[2mm]
 \mbox{set (d):} && \!\!\!\!\!
                     \frac{1}{4}\,\Gamma_{\!\!ff_1\!V_1}\,
                     \Gamma_{\!\!f_1f_2V_3}\,\Gamma_{\!\!ff_2V_2}\,G_{132}\,
                     \biggl\{ \Bigl[ \cK^{\,(1)}(s,m_f^2,M_1,y_1,z_1) 
                             + \cK^{\,(1)}(s,m_f^2,M_2,y_1,z_1) \Bigr]
                     \times 
                     \no\\[1mm]
                  && \!\!\!\!\!\hphantom{\frac{1}{4}\,\Gamma_{\!\!ff_1\!V_1}\,
                               \Gamma_{\!\!f_1f_2V_3}}
                     \times\,\,\cK^{\,(1)}(s,m_f^2,M_3,y_2,z_2)\,
                     \Theta(y_2-y_1)\,\Bigl[ 1 + 3\,\Theta(z_1-z_2) \Bigr]
                     \no\\[3mm]
                  && \!\!\!\!\!\hphantom{\frac{1}{4}\,\Gamma_{\!\!ff_1\!V_1}`\,
                               \Gamma_{\!\!f_1f_2V_3}\,\Gamma_{\!\!ff_2V_2}\,
                               G_{132}\,A}
                     {}+\cK^{\,(1)}(s,m_f^2,M_1,y_1,z_1)\,
                        \cK^{\,(1)}(s,m_f^2,M_2,y_2,z_2)\,
                     \times 
                     \no\\[2mm]
                  && \!\!\!\!\!\hphantom{\frac{1}{4}\,\Gamma_{\!\!ff_1\!V_1}\,
                               \Gamma_{\!\!f_1f_2V_3}}
                     \times\,\,\Bigl[ 3 + \Theta(y_1-y_2)\,\Theta(z_2-z_1)
                                    + \Theta(y_2-y_1)\,\Theta(z_1-z_2) \Bigr]
                     \biggr\}~,
                     \no\\[2mm]
  \mbox{set (e):} && \!\!\!\!\!
                     -\,\frac{1}{2}\, \Gamma_{\!\!ff_1\!V_1}\,\Gamma_{\!\!ff_1\!V_2}\,
                     G_{134}\,G_{234}\,\Bigl[ 
                     \cK^{\,(1)}(s,m_f^2,M_3,y_2,z_2) 
                     + \cK^{\,(1)}(s,m_f^2,M_4,y_2,z_2) \Bigr] 
                     \times 
                     \no\\[2mm]
                  &&  \!\!\!\!\! \times\,\, \Bigl[ \cK^{\,(1)}(s,m_f^2,M_1,y_1,z_1) +  
                                          \cK^{\,(1)}(s,m_f^2,M_2,y_1,z_1) \Bigr]\,
                     \Theta(y_2-y_1)\,\Theta(z_1-z_2)~.                      
\ea  
The totally antisymmetric coupling
$\,e\,G_{ijl}\,$ is the triple gauge-boson coupling with all three gauge-boson
lines ($i,j,l$) defined to be incoming at the interaction vertex. In our 
convention this coupling is fixed according to $G_{\ga W^+ W^-}=1\,$ and 
$G_{Z W^+ W^-}=-\cos\theta_{\rm w}/ \sin\theta_{\rm w}$. In the calculation
    of the integration kernel of set (e) we used the following couplings
    between the would-be Goldstone boson $\phi$ and two gauge bosons: $
    e\,C_{\phi W \gamma}= -e \,M_{_W}\,$ and $\,e\,C_{\phi W Z}= -e \,M_{_W} \sin
    \theta_{\rm w}/\cos\theta_{\rm w} $. The $\,W^{\pm}\! - \phi^{\pm}$ mixing
    propagators, entering the same calculation, are given by 
\bea
M_{\mu}^{\phi^{\pm} W^{\pm}}(k,n) = M_{\mu}^{W^{\pm}  \phi^{\pm}}(k,n) =
    \frac{\mp i M_{_W}}{k^2-M_{_W}^2+i\ep}\, \frac{k^0 n_{\mu}}{\vec{k}^{\,2}}~,
\ea
with $k$ the momentum of the $W^{\pm}$-boson.\\

\ni
Note that several of
the integration kernels involve a specific ordering in the energy variables 
$z_i$ [in set (e) and part of set (d)] and/or
 the angular
variables $y_i$ [in sets (a),(e) and part of set (d)]. The angular ordering in set (a), for
instance, is caused by the fact that in the outer ($k_1$) loop-integral the 
incoming fermion momentum is $\,p$, yielding $\,(p-k_1)^2 \approx -s\,y_1$, 
whereas in the inner ($k_2$) loop-integral it is $\,p-k_1$, yielding 
$\,(p-k_1-k_2)^2 \approx -s\,(y_1+y_2)$. Therefore, only for $\,y_2\gg y_1\,$ 
a large logarithm develops. The energy ordering occurs, for instance, when the 
soft gauge-boson momentum of the outer loop-integral is the incoming momentum 
of the inner loop-integral [see set (e)]. The Sudakov parametrisation for 
$\,k_2\,$ is based on $\,k_1\,$ in that case and the large logarithms only 
develop if $\,k_2^0 \ll k_1^0$. The multitude of terms contributing to 
the integration kernels of sets (d) and (e) originate from the various soft
gauge-boson regimes that are possible in those diagrams. \\

\ni Note that certain diagrams look possible at first sight, but are in fact 
forbidden as a result of the charged current interactions of the $W$ bosons. 
For instance, in set (b) it is not possible to exchange two $W$ bosons without 
reversing the fermion-number flow (given by the direction of the Dirac 
propagator lines). Adding up all possible contributions, we find for the full 
two-loop Sudakov correction factor for right- and left-handed 
fermions/antifermions (see Sect.~\ref{sec:integrals})
\begin{subequations} 
\label{delta2}
\bea
\label{delta2+} 
 \delta^{\,(2)}_{f_R} &=& \delta^{\,(2)}_{\bar{f}_L} \ =\ \frac{1}{2}\,
        \left( \delta^{\,(1)}_{f_R}\right)^2  \\
\label{delta2-}
 \delta^{\,(2)}_{f_L} &=& \delta^{\,(2)}_{\bar{f}_R} \ =\ \frac{1}{2}\,
        \left( \delta^{\,(1)}_{f_L}\right)^2~.
\ea
\end{subequations}

\ni
{}From Eqs.~(\ref{delta2+}) and (\ref{delta2-}) we deduce our main statement,
namely that the virtual electroweak two-loop Sudakov correction factor is
obtained by a mere exponentiation of the one-loop Sudakov correction
factor. This is in agreement with the corresponding assumptions in Ref.~\cite{melles}. We also note that, 
in adding up all the contributions, we find that the 
`rainbow' diagrams of set (a) yield the usual exponentiating terms plus an 
extra term for left-handed fermions similar to the one found in
Ref.~\cite{paolo2}. This extra term originates from 
the charged-current interactions and is only non-vanishing as a result of the
mass gap between the massless photon and the massive $Z$ boson. We therefore disagree with the statement in
Ref.~\cite{melles} that only rainbow diagrams contribute in physical gauges
like the axial or the Coulomb gauge. Whereas in Ref.~\cite{paolo2} the extra 
term was interpreted as a source of non-exponentiation, we observe that it 
in fact cancels against a specific term originating from the triple gauge-boson
diagrams of set (d). Similar (gauge) cancellations take place between the 
`crossed rainbow' diagrams of set (b), the reducible diagrams of set (c), and 
another part of the triple gauge-boson diagrams of set (d). Finally,  
the left-over terms of set (d) get cancelled by the contributions from the
gauge-boson self-energy (`frog') diagrams of set (e). Hence,  the
cancellations that take place automatically in unbroken gauge theories also
hold in the SM, in spite of it being a theory with more than one
scale in the sense that the on-shell poles for
photons and $Z$ bosons do not coincide and therefore lead to different
 on-shell residues (see Sect.~\ref{sec:integrals}).\\

\ni Comparing with the study in Ref.~\cite{kuehn1}, we can make the following 
remark. A treatment of pure weak gauge-boson effects without 
reference to the photonic interactions breaks gauge-invariance, since the 
photon has an explicit $SU(2)$ component. This holds even if the photon is 
treated fully inclusively as in Ref.~\cite{kuehn1}. Such a separation would 
require a very careful definition, for instance in terms of the typical energy 
regimes that govern the Sudakov effects of pure electromagnetic origin 
(ultrasoft energies: $\la/\sqrt{s} \le z \le M/\sqrt{s}$) and collective
electroweak origin (soft energies: $M/\sqrt{s} \le z \ll 1$).

\subsection{Some useful integrals}
\label{sec:integrals}

In order to make our analysis more accessible, we present in this subsection 
the relevant loop-integrals, using the generic notation
\bea 
  I^{(i)} &=& \Bigl( -\,\frac{\alpha}{\pi} \Bigr)^i
              \int_0^1 \frac{\dd y_1}{y_1} \,\int_{y_1}^1 \frac{\dd z_1}{z_1}\,
              \ldots            
              \int_0^1 \frac{\dd y_i}{y_i} \,\int_{y_i}^1 \frac{\dd z_i}{z_i}\,
              \cK^{(i)}(y_1,z_1,\ldots,y_i,z_i)~.
\ea
At one-loop level two types of kernels occur:
\bea
\label{one-loop}
  \cK^{\,(1)}(s,m_f^2,M,y_1,z_1)   &:& I^{(1)} = {\rm L}(M,M)~, \no\\[1mm]
  \cK^{\,(1)}(s,m_f^2,\la,y_1,z_1) &:& I^{(1)} = {\rm L_{\ga}}(\la,m_f)~.
\ea
The functions $\,{\rm L}(M_1,M_2)\,$ and $\,{\rm L_{\ga}}(\la,M_1)\,$ are the 
ones defined in Eqs.~(\ref{L}) and (\ref{Lgamma}). At two-loop level seven new 
types of kernels occur. Four have angular ordering:
\bea
\label{two-loop/ang}
  \cK^{\,(1)}(s,m_f^2,M,y_1,z_1)\,\cK^{\,(1)}(s,m_f^2,M,y_2,z_2)\,
  \Theta(y_2-y_1) 
       &:& I^{(2)} = \frac{1}{2}\,{\rm L}^2(M,M)~, \no\\[1mm]
  \cK^{\,(1)}(s,m_f^2,\la,y_1,z_1)\,\cK^{\,(1)}(s,m_f^2,\la,y_2,z_2)\,
  \Theta(y_2-y_1) 
       &:& I^{(2)} = \frac{1}{2}\,{\rm L_{\ga}}^{\!\!2}(\la,m_f)~, \no\\[1mm]
  \cK^{\,(1)}(s,m_f^2,M,y_1,z_1)\,\cK^{\,(1)}(s,m_f^2,\la,y_2,z_2)\,
  \Theta(y_2-y_1) 
       &:& I^{(2)} = \frac{7}{12}\,{\rm L}^2(M,M)~, \no\\[2mm]
  \cK^{\,(1)}(s,m_f^2,\la,y_1,z_1)\,\cK^{\,(1)}(s,m_f^2,M,y_2,z_2)\,
  \Theta(y_2-y_1) 
       &:& I^{(2)} = {\rm L}(M,M)\,{\rm L_{\ga}}(\la,m_f) \no\\[1mm]
       & & \hphantom{I^{(2)} =} - \frac{7}{12}\,{\rm L}^2(M,M)~, 
\ea
and three have double ordering in both angle and energy:
\bea
\label{two-loop/en}
  \cK^{\,(1)}(s,m_f^2,M,y_1,z_1)\,\cK^{\,(1)}(s,m_f^2,M,y_2,z_2)\,
  \Theta(y_2-y_1)\,\Theta(z_1-z_2)
       &:& I^{(2)} = \frac{1}{4}\,{\rm L}^2(M,M)~, \no\\[1mm]
  \cK^{\,(1)}(s,m_f^2,M,y_1,z_1)\,\cK^{\,(1)}(s,m_f^2,\la,y_2,z_2)\,
  \Theta(y_2-y_1)\,\Theta(z_1-z_2)
       &:& I^{(2)} = \frac{1}{3}\,{\rm L}^2(M,M)~, \no\\[1mm]
  \cK^{\,(1)}(s,m_f^2,\la,y_1,z_1)\,\cK^{\,(1)}(s,m_f^2,M,y_2,z_2)\,
  \Theta(y_2-y_1)\,\Theta(z_1-z_2)
       &:& I^{(2)} = \frac{2}{3}\,{\rm L}(M,M)\,{\rm L}(M,m_f) \no\\[1mm]
       & & \hphantom{I^{(2)} =} {}-\frac{1}{4}\,{\rm L}^2(M,M)~.
\ea
Note that in the case of double ordering the collinear cut-off $m_f^2$ of the
$y_2$ integral is in fact redundant.

\section{Conclusions} 

In order to settle the controversy in the literature concerning the 
resummation of virtual Sudakov logarithms in the process $e^+e^- \to f\bar{f}$,
we have calculated these Sudakov logarithms at one- and two-loop level in the 
Coulomb gauge. In this special gauge all the relevant contributions, involving
the exchange of collinear-soft gauge bosons, are contained in the self-energies
of the external on-shell particles.\\

\ni Our one-loop results are in agreement with the calculations
in the literature, including the distinctive terms
originating from the mass gap between the photon and the weak gauge bosons.  At two-loop level our findings are consistent with an 
exponentiation of the one-loop results, in disagreement with claims in the
literature that non-exponentiating terms should emerge at two-loop level. 
Based on these results and a similar study for transverse/longitudinal gauge-boson
production processes, we conclude that as far as the balance between the one- and two-loop virtual 
Sudakov logarithms are concerned,  the SM  behaves like an unbroken theory at 
high energies. This conclusion can be extended to real-emission processes in
a relatively straightforward way. After all,
since the Sudakov logarithms originate from the exchange of soft, effectively 
on-shell gauge bosons, many of the features derived for the virtual corrections
will be intimately related to properties of the corresponding real-emission 
processes.

\end{document}